\documentclass[11pt,twoside]{article}
\usepackage{asp2010}

\begin{document}

\title{AGN Winds and the Black-Hole - Galaxy Connection}

\author{Kastytis Zubovas$^{1}$, Andrew R. King$^1$
  \affil{$^1$ Dept. of Physics \& Astronomy, University of Leicester,
    Leicester, LE1 7RH, UK; mailto:ark@astro.le.ac.uk}}

\begin{abstract}

During the last decade, wide--angle powerful outflows from AGN, both on parsec
and kpc scales, have been detected in many galaxies. These outflows are widely
suspected to be responsible for sweeping galaxies clear of their gas. We
present the analytical model describing the propagation of such outflows and
calculate their observable properties. Large--scale AGN--driven outflows
should have kinetic luminosities $\sim \eta L_{\rm Edd}/2 \sim 0.05L_{\rm
  Edd}$ and momentum rates $\sim 20L_{\rm Edd}/c$, where $L_{\rm Edd}$ is the
Eddington luminosity of the central black hole and $\eta\sim 0.1$ its
radiative accretion efficiency. This creates an expanding two--phase medium in
which molecular species coexist with hot gas, which can persist after the
central AGN has switched off. This picture predicts outflow velocities $\sim
1000 - 1500$~km\,s$^{-1}$ and mass outflow rates up to $4000~\rm M_{\odot}\,{\rm
  yr}^{-1}$ on kpc scales, fixed mainly by the host galaxy velocity dispersion
(or equivalently black hole mass). We compare our prediction with recent
observational data, finding excellent agreement, and suggest future
observational tests of this picture.

\end{abstract}

\section{Introduction}

Recently, two sets of observations have allowed us to gain a better
understanding of the interaction between AGNs and their host galaxies. These
are observations of high--velocity wide--angle winds emanating from the
vicinity of the SMBH, which have been detected in a large fraction of AGNs
\citep{Tombesi2010A&A, Tombesi2010ApJ}; and detection of kpc--scale
quasi--spherical outflows in active galaxies, with enough power and mass flow
to sweep their host galaxies clear of gas \citep{Feruglio2010A&A,
  Rupke2011ApJ, Sturm2011ApJ, Riffel2011MNRASb, Riffel2011MNRAS}. These
outflows have kinetic power equal to a few percent of the Eddington luminosity
of the central black hole and their momentum flow rate is approximately an
order of magnitude greater than $L_{\rm Edd}/c$.

In this paper, we show how the two types of flows can be explained within the
framework of AGN wind feedback. Radiation pressure from an accreting SMBH
expels gas in form of a wind from the nucleus \citep[e.g.][]{Pounds2003MNRASb,
  Pounds2003MNRASa}, which then pushes the ambient gas in the host galaxy and
produces an outflow. In recent work \citep{King2011MNRAS,Zubovas2012arXiv} we
have shown that large--scale energy--driven flows (see Section
\ref{sec:outflow}) can indeed drive much of the interstellar gas out of a
galaxy bulge on a dynamical timescale $\sim 10^8$~yr, leaving it red and dead.
The remaining mass of the bulge is then similar to the value set by the
observed black--hole -- bulge--mass relation
\citep[e.g.][]{Haering2004ApJ}. The observable features of such outflows --
velocities, kinetic powers and mass and momentum flow rates -- are consistent
with observations. Therefore AGN outflows appear capable of sweeping galaxies
clear of gas.

\section{Close to the SMBH -- winds}

Radiation pressure from an AGN accreting at close to its Eddington limit can
expel gas from the vicinity of the nucleus with a momentum rate
\begin{equation}\label{eq:mom}
\dot{M}_{\rm w} v_{\rm w} = \frac{L_{\rm Edd}}{c},
\end{equation}
as the wind on average has scattering optical depth $\sim 1$ and absorbs all
of the radiation momentum. This creates a mildly relativistic diffuse wind
\citep[$\dot{M}_{\rm w} \sim \dot{M}_{\rm Edd}$ and $v_{\rm w} \sim \eta c
  \sim 0.1c$, where $\eta \simeq 0.1$ is the accretion radiative
  efficiency][]{King2003ApJ,King2003MNRASb}. Observations of blueshifted
X--ray iron absorption lines corresponding to velocities $\sim 0.1c$
\citep[e.g.][]{Pounds2003MNRASb, Pounds2003MNRASa, Tombesi2010A&A,
  Tombesi2010ApJ} reveal that the majority of quasars produce such winds.  The
winds have momentum and energy rates
\begin{equation}\label{eq:pdotedot}
\dot{P}_{\rm w} \sim \frac{L_{\rm Edd}}{c}; \; \; \; \dot{E}_{\rm w} =
\frac{1}{2} \dot{M}_{\rm w} v_{\rm w}^2 \sim 0.05 L_{\rm Edd}.
\end{equation}

\section{Out in the galaxy -- outflows} \label{sec:outflow}

It is clear that the wind has enough kinetic power to drive the observed large
scale outflow, provided that it can efficiently transfer this power to the
ISM. In order for this to happen, two conditions must be satified. First, most
of the sightlines from the SMBH must be covered with diffuse medium. Second,
the wind cannot cool efficiently. As the wind hits the ISM, it shocks and
heats to $T \sim 10^{11}$~K. At this temperature, the most efficient cooling
process is inverse Compton scattering of the photons in the AGN radiation
field \citep{Ciotti1997ApJ}. The efficiency of this process drops with
increasing shock radius, thus the cooling timescale increases as $R^2$. Since
the outflow velocity does not depend strongly on radius
\citep{King2010MNRASa,King2011MNRAS}, the flow timescale only increases as
$R$. Therefore, there is a critical radius, $R_{\rm cool} \sim 1$~kpc, within
which the shock can be cooled efficiently, whereas outside most of its energy
is retained and transferred to the outflow. The two cases are called
momentum--driven and energy--driven flows, respectively; their salient
features are shown schematically in Figure \ref{fig:outflow}.

\subsection{Momentum--driven outflow}

An efficiently cooled shocked wind gas is compressed to high density and
radiates away almost all of its original kinetic energy, retaining and
communicating only its pressure, which is equal to the pre--shock ram pressure
$\dot P_{\rm w} \simeq L_{\rm Edd}/c \propto M$, to the host ISM.

For an isothermal ISM density distribution with velocity dispersion $\sigma$
and gas fraction $f_c$ (the ratio of gas density to background potential
density) the behaviour of the flow depends on the black hole mass $M$
\citep{King2003ApJ, King2010MNRASa}. For $M < M_{\sigma}$, where
\begin{equation} \label{msig}
M_{\sigma} = {f_c\kappa\over \pi G^2}\sigma^4 \simeq 4\times
10^8\rm M_{\odot}\sigma_{200}^4,
\end{equation}
with $f_c = 0.16$ and $\sigma_{200} = \sigma/(200~{\rm km\,s^{-1}})$, the wind
momentum is too weak to drive away the swept--up ISM, and the flow stalls. For
$M > M_{\sigma}$ the wind drives the swept--up ISM far from the
nucleus, quenching its own gas supply and further accretion. Therefore,
$M_\sigma$ represents an approximate upper limit to the SMBH mass distribution
\citep[see][for more details]{Power2011MNRAS}. The calculated mass is very
similar to that obtained from observations of the $M-\sigma$ relation, despite
having no free parameter.

\subsection{Energy--driven outflow}

\begin{figure}
  \plotone{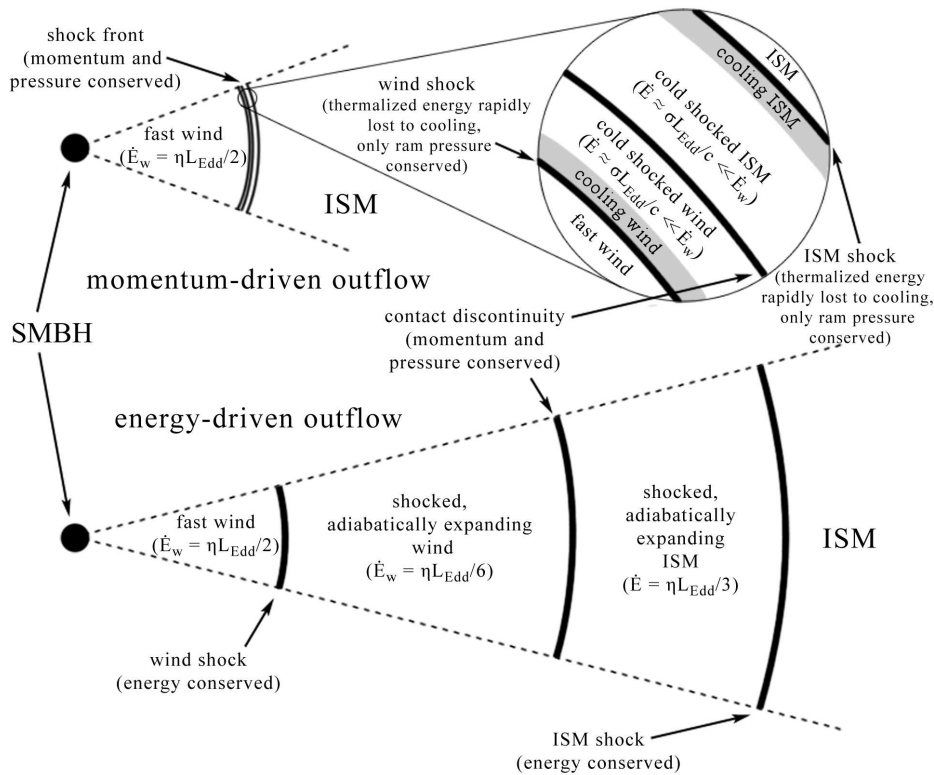}
  \caption{Schematic picture of AGN outflows. A wind with $v_{\rm w} \sim
    0.1c$) impacts the ISM of the host galaxy, producing a shock on either
    side of the contact discontinuity. Within $\sim 1$~kpc of the nucleus
    (top), the shocks cool rapidly and radiate away most of their energy,
    leading to outflow kinetic energy $\sim (\sigma/c)L_{\rm Edd}$. In an
    energy--driven outflow (bottom), the shocked regions expand adiabatically,
    communicating most of the kinetic energy of the wind to the outflow, which
    is then able to sweep the galaxy clear of gas.}
  \label{fig:outflow}
\end{figure}

A large--scale ($\gtrsim 1$~kpc) outflow becomes energy driven. It is
essentially adiabatic, and has the wind energy rate, i.e. $\dot E_{\rm out}
\simeq \dot E_{\rm w} \sim 0.05L_{\rm Edd}$ (from Equation \ref{eq:pdotedot}). The hot
bubble's thermal expansion makes the driving into the host ISM more vigorous
than in the momentum--driven case.  Observed galaxy--wide molecular outflows
must be energy--driven, as demonstrated directly by their kinetic energy
content (cf. Equation \ref{eq:pdotedot}). The adiabatic expansion of the
shocked wind pushes the swept--up interstellar medium in a `snowplow'. In
\citet{King2011MNRAS} we derive the analytic solution for the expansion of the
shocked wind in a galaxy bulge with an isothermal mass distribution. With AGN
luminosity $lL_{\rm Edd}$, all such solutions tend to an attractor
\begin{equation}
\dot R = v_e \simeq \left[\frac{2\eta lf_c}{3f_g}\sigma^2c\right]^{1/3} \simeq
925\sigma_{200}^{2/3}(lf_{\rm c}/f_{\rm g})^{1/3}~{\rm km\ s}^{-1}
\label{ve}
\end{equation} 
until the AGN switches off when the shock is at some radius $R =
R_0$. Subsequently, the expansion speed decays with $x = R/R_0\geq 1$ as
\begin{equation}
\dot R^2 = 3\biggl(v_e^2 + {10\over 3}\sigma^2\biggr)\biggl({1\over
  x^2} - {2\over 3x^3}\biggr) - {10\over 3}\sigma^2.
\label{dotr}
\end{equation}
In Eq. (\ref{ve}), the current gas fraction $f_g$ may be lower than $f_c$
(cf. Eq. \ref{msig}). The outflow persists for an order of magnitude longer
than the duration of the quasar outburst that is driving it, and reaches radii
of $10^4 - 10^5$~pc. It is evident that energy--driven outflows are capable of
sweeping gas out of galaxies, quenching further star formation and
establishing the SMBH -- bulge mass relationship \citep{Power2011MNRAS}.

\subsection{Observable outflow parameters}

The solutions (\ref{ve}, \ref{dotr}) describe the motion of the contact
discontinuity see Figure \ref{fig:outflow}). Outflows are usually observed in
molecular gas, which is embedded in the outflowing shell \citep[see][for more
  details]{Zubovas2012arXiv}, which moves with velocity
\begin{equation}
v_{\rm out} = {\gamma + 1\over 2}\dot R \simeq
1230\sigma_{200}^{2/3}\left({lf_c\over f_g}\right)^{1/3}~{\rm km\ s}^{-1}
\label{vout}
\end{equation}
from adiabatic shock conditions, using $\gamma= 5/3$, and the mass outflow
rate is
\begin{equation} 
\dot{M}_{\rm out} = \frac{{\rm d}M(R_{\rm out})}{{\rm d}t} = {(\gamma +
  1)f_{\rm g} \sigma^2\over G}\dot R = \frac{\eta(\gamma +
  1)}{4}\frac{f_g}{f_c}\frac{\dot Rc}{\sigma^2}\dot{M}_{\rm Edd},
\end{equation} 
assuming $M = M_{\sigma}$. If the AGN luminosity is still close to Eddington
and $f_{\rm g} = f_{\rm c}$, the mass loading factor ($f_{\rm L} \equiv
\dot{M}_{\rm out} / \dot{M}_{\rm Edd}$) and mass outflow rates are
\begin{equation}
f_{\rm L} = \left({2\eta c\over 3\sigma}\right)^{4/3}\left({f_g\over
  f_c}\right)^{2/3}{l^{1/3}\over \dot m} \simeq
460\sigma_{200}^{-4/3}{l^{1/3}\over \dot m}; \;\;\; \dot{M}_{\rm out} \simeq
3700\sigma_{200}^{8/3}l^{1/3}~\rm M_{\odot}\,{\rm yr}^{-1}.
\label{eq:flmout}
\end{equation}
If the central quasar is no longer active, $\dot{M}_{\rm out}$ is lower by
$\dot R/v_e$, with $\dot R$ given by Eq. (\ref{dotr}).

One can show from Equations (\ref{vout}) and (\ref{eq:flmout}) that
$\dot{M}_{\rm out}v_{\rm out}^2/2 \simeq 0.05 L_{\rm Edd}$, i.e. most of the
wind kinetic energy is transferred to the outflow, as expected for energy
driving (more precisely, while the quasar is active, the outflow contains
$2/3$rds of the total energy). We can also derive an expression for the
momentum flow rate $\dot{P}$ in the outflow:
\begin{equation} 
\dot{P}_{\rm out} = \frac{L_{\rm Edd}}{c} f_{\rm L}^{1/2} \sim 20
\sigma_{200}^{-2/3} l^{1/6} \frac{L_{\rm Edd}}{c}.
\end{equation}

\begin{table*}
\centering

  \caption{Outflow parameters: observation versus prediction for a sample of AGN}

  \setlength{\extrarowheight}{1.5pt}

  \begin{tabular}{c | c c | c c c | c c c }
    \hline \hline Object & $\dot{M}_{\rm out}$ & $v_{\rm out}$ &
    $\frac{\dot{E}_{\rm out}}{0.05 L_{\rm bol}}$ & $\frac{\dot M_{\rm out} v_{\rm out}
      c}{L_{\rm bol}}$ & $f_{\rm L}$ & $\dot{M}_{\rm pred.}$ & $v_{\rm pred.}$
    & $f_{\rm L, pred.}$ \\

    \hline 

    Mrk231$^{(a)}$ & $420$ & $1100$ & $0.66$ & $18$ & $490$ & $880$ &
    $810$ & $840$ \\
                                                                                                                      
    Mrk231$^{(b)}$ & $700$ & $750$ & $0.51$ & $20$ & $820$ & $880$ &
    $810$ & $840$ \\
                                                                                                                      
    Mrk231$^{(c)}$ & $1200$ & $1200$ & $1.0$ & $25$ & $1400$ & $1150$ &
    $1060$ & $1110$ \\
                                                                                                                      
    IRAS 08572+3915$^{(c)}$ & $970$ & $1260$ & $2.1$ & $50$ & $1200$ &
    $950$ & $875$ & $910$ \\
                                                                                                                      
    IRAS 13120--5453$^{(c)}$ & $130$ & $860$ & $0.88$ & $31$ & $1080$ &
    $220$ & $610$ & $1870$ \\
                                                                                                                      
    \hline
    \hline
  \end{tabular}

\begin{list}{}{}
\item[ ] {\footnotesize First two columns: observed mass flow rate (in $\rm
  M_{\odot}$~yr$^{-1}$) and velocity (in km s$^{-1}$) of large--scale outflows
  in molecular (Mrk231, IRAS 08572+3915 and IRAS 13120--5453) and warm ionised
  gas (Mrk1157). Middle three columns: quantities derived from
  observations. Last three columns: mass flow rate, velocity and mass loading
  factor derived from our equations (\ref{vout}) and (\ref{eq:flmout}). All
  derived quantities show good agreement with those observed and with each
  other.

  References: $^a$ - \citet{Rupke2011ApJ}; $^b$ - \citet{Feruglio2010A&A};
  $^c$ - \citet{Sturm2011ApJ}.}
\end{list}

  \label{table:obs}
\end{table*}

\section{Discussion} \label{sec:discuss}

We see that in principle, large--scale wide--angle outflows driven by a mildly
relativistic wind launched by the AGN radiation pressure can sweep galaxies
clear of gas. The observable properties of such outflows are typical
velocities $v_{\rm out} \sim 1000 - 1500$~km\,s$^{-1}$ and mass flow rates up
to $\dot M_{\rm out} \sim 4000~\rm M_{\odot}\,{\rm yr}^{-1}$ (Equations
(\ref{vout}) and(\ref{eq:flmout})). The outflows should have mechanical
luminosities $\dot E_{\rm out} \sim (\eta/2) L_{\rm Edd} \sim 0.05 L_{\rm
  Edd}$, but (scalar) momentum rates $\dot P _{\rm out}\sim 20 L_{\rm Edd}/c$,
consistent with observations (see Table \ref{table:obs}).

Such outflows leave several observable signatures. Cold gas clumps entrained
within the shell produce the observed molecular emission. The inner wind shock
accelerates cosmic ray particles, which can emit synchrotron radiation in the
radio band and produce gamma rays when interacting with the ISM. These
signatures resemble those of the gamma--ray emitting bubbles in our Galaxy
recently discovered by {\it Fermi} \citep{Su2010ApJ}, which can be explained
as relics of a short quasar outburst about 6~Myr ago \citep[][also the
  contribution by Zubovas to this volume]{Zubovas2011MNRAS}.

\acknowledgments
 
We thank the conference organizers for their hospitality. Research in
theoretical astrophysics at Leicester is supported by an STFC Rolling
Grant. KZ is supported by an STFC research studentship.


\end{document}